# Elephant Flows Detection Using Deep Neural Network, Convolutional Neural Network, Long Short Term Memory and Autoencoder

Getahun Wassie Geremew
IP Networking and Mobile Internet Track
Addis Ababa University
Addis Ababa, Ethiopia
getahunws12@gmail.com
getahun.wassie@psi.gov.et

Jianguo Ding
Department of Computer Science
Blekinge Institute of Technology
371 79 Karlskrona, Sweden
jianguo.ding@bth.se

**Abstract**

Currently, the wide spreading of real-time applications such as VoIP and videos-based applications require more data rates and reduced latency to ensure better quality of service (QoS). A well-designed traffic classification mechanism plays a major role for good QoS provision and network security verification. Port-based approaches and deep packet inspections (DPI) techniques have been used to classify and analyze network traffic flows. However, none of these methods can cope with the rapid growth of network traffic due to the increasing number of Internet users and the growth of real time applications. As a result, these methods lead to network congestion, resulting in packet loss, delay and inadequate QoS delivery. Recently, a deep learning approach has been explored to address the time-consumption and impracticality gaps of the above methods and maintain existing and future traffics of real-time applications.

The aim of this research is then to design a dynamic traffic classifier that can detect elephant flows to prevent network congestion. Thus, we are motivated to provide efficient bandwidth and fast transmission requirements to many Internet users using SDN capability and the potential of Deep Learning.

Specifically, DNN, CNN, LSTM and Deep autoencoder are used to build elephant detection models that achieve an average accuracy of 99.12%, 98.17%, and 98.78%, respectively. Deep autoencoder is also one of the promising algorithms that does not require human class labeler. It achieves an accuracy of 97.95% with a loss of 0.13. Since the loss value is closer to zero, the performance of the model is good.

Therefore, the study has a great importance to Internet service providers, Internet subscribers, as well as for future researchers in this area.

Keywords: SDN, QoS, Elephant Flow, DNN, CNN, LSTM, Autoencoder

1. Introduction

Nowadays, Internet technology turning to real-time applications that requires high bit rates and strict delay for better quality of service (QoS) provision [1]. Real time applications such as Voice over Internet Protocol (VoIP), video conferencing, online gaming [2] [3], online transactions [4] and virtual online classroom [5] become hot research areas for real tme applcations due to the rapid growth of user interest in audio and video and the availability of integrated systems that can deliver multimedia data at a lower cost [6] [7]. These benefits have been achieved through the establishment of many multimedia institutions such as Google, Akamai, level 3, limelight and Kankan [6] [8]. However, real time applications demand a reliable and high-speed data (high startup and playback speed), characterized as large streams(elephants flows) or small streams (mouse flows), with strinent QoS and QoE requirements.

Elephant flows are less in numberous (10%) and exhibit long-lived flows that potentially fill network buffers end-to-end [9]. Elephant flows cause node congestion , delays, and packet loss if it is not managed appropriately [10]. To minimize or avoid this gap, our work aims to develop a dynamic traffic model to achieve a more predictable network and QoS in Software-Defined Networks (SDN) network by using deep learning approach. The proposed model aims to detect elephant flows to minimize network constraints such as latency, packet loss and controller congesion. Several mechanisms have already been proposed to detect elephant flows [9]. However, these mechanisms do not provide a general and standardized method for elephant detection. Various fixed-sized thresholds have been explored to classify flows in openswitches and controllers, which can lead to a high rates of false positives and false negatives [11]. Thus, there is a need to find standardized, dybamic and optimal thresholds that can consider flow size, duration, packet size and application type as heuristics i.e.,If these parameters are used as heuristics, the network extensible and satisfies various QoS requirements. As IP networks combine the text, voice and video data, dynamic flow classification is required to categorize traffic of different and future applications and provide the required services [6].

Traffic classification can be based on protocol types (e.g., UDP, TCP, FTP or HTTP), application type (e.g., Skype, Chat or Torrent) and traffic types (e.g., browsing, downloading or video chat) [12] [13]. However, classifying traffic at application level becomes difficult because there are more than thousands of applications and new application are always being developed. Therefore, it is better to track the traffic flows at network layer by classifying as elephant and mice to achieve efficient QoS resource allocation. Therefore,developing a good classifier is one of the prerequisites for providing appropriate and adequate QoS and QoE in advanced traffic management of real time applications [14]. The traffic classifier can be defined according to the classification objective i.e., typical objective of this work is to provide efficient QoS and QoE provisioning [15].

Traffic classifcation is a key task for any Internet Service Providers (ISPs) or network administrators QoS provision [19]. For this reason, it have been experimented using deep packet inception (DPI) [16] and machine learning approaches [17] [18]. To provide better Internet services according to application requirements, machine learning approaches are more recognized than DPI and of architectures implementation [20]. Deep learning is the state of the art machine learning approach [21] that inspired us to develop our proposed stream classification model [22] [21]. Specifically, we developed an elephant flow detection model for SDN using deep neural network (DNN), convolutional neural network (CNN), long short term memory (LSTM) and autoencoder algorithms. These algorithms have a dropout function to remove unnecessarily information [23]. Moreover, they overcome the problem of overfitting and underfitting during model training by weight regularization(Adam),optimal epoch and batch normalization techniques [24].

DNN is a multi-layer perceptron (MLP) types of neural network that has input layer, hidden layer and output layer in one forward direction without going backward trsnmission [25]. The Inhanced NN,DNN, is required to manage complex raffic from voice and video services. The quality of multimedia services on the Internet depends of congestions, failures and other anomalies in the network. Therofre, we need a more advanced ways to prevent these problems. Furthermore, DNN leads to success for internet service providers QoS guarantee the quality of service. QoS measuring and resource prediction is possible with DNN for of distributed multimedia applications. DNN can forecast future traffic variations as accurately as possible to predict the user network behavior.

CNN is a feature extractor used to filter and preprocess data by defining local correlation between network neurons of neighboring layers to provide abstract representations of the input features. The loss function renders feedback signal for the learning purpose, and the optimizer is used to determine how learning proceeds [21] [26].

LSTM is an extension of RNN that provides the capability of "long term memory" in addition to short-term memory. It stores a list of all of the previous information in its memory and makes it available for training current neural neuron [27] [28].

Deep autoencoder uses an input layer, bottleneck and an output layer. It extracts features during training and it usually works with CNN. CNN-based autoencoder uses CNN to filter content bearing words, numbers or images values from the given input Instance (record) [26].

The proposed classifier classifies traffics into elephant and mouse flows as the current traffic flow management is time-consuming and impractical to update the list of existing and future applications to transmit their heavy loaded traffic. Once we identify elephant flows, it is possible to implement a clustering model in SDN controller for route assignment of elephant flows based on elephant size to avoid overloaded of links in a networks.

In this work, research questions are prepared to answer the existing network QoS constarints. The questions include "By how much DNN, CNN, LSTM and autoencoder detect elephant flows from mice flows for good QoS?". The performance of these algorrthms also seen in terms of accuracy and execution time over different QoS based datasets. Optimal epoch iteration was experimented to overcome under fit and over fit challenges. The effect of batch size was also investigated during flow classification model development at constant and optimal epoch.

After presenting th the introduction in this section I, Realated work is discussed in Section II, We presented our proposed traffic model architecture and modeling concepts in section III. In section IV, experimentation results are discussed and evaluated. Finally, we presented the conclusion and future works in Section V.

## 2. Related Work

The ever-growing number of applications with their huge and heterogeneous traffic needs more advanced traffic management mechanisms to achieve good end-to-end quality on the Internet [29]. In particular, real-time online applications (audio/video) require dynamic traffic management to prevent the negative effect of elephant flows in the network at runtime [3] [30].

Multimedia streaming, real-time interactive applications and parallel processing in data centers are some of the examples of applications that require QoS guarantees for an enhanced quality of experience (QoE) to users [2]. The need of good traffic managment in real time applications and QoS requirements has motivated many research works in network field. The efforts resulted in several proposals with a number of research approaches. Research approaches that have been proposed for real-time traffic management include architecture-based [31] [2] [32], port-based [13] [33], payload-based [34], machine learning-based [10] [22] [12] [11] [30] and deep learning-based [21] [24] [35] [36] [37] [38] [39] approaches. However, the port-based and payload-based options are not efficient in detecting signatures from payloads after encryption[40]. Therefore, QoS cannot be achieved with these approaches. To achieve good QoS, authors such as Alexandre and et al. [2] focused on architecture-based solutions.

Alexander and et al [2] proposed an SDN-based architecture that provides QoS on distributed applications. SDN constitutes an emerging paradigm that facilitates the creation and introduction of new abstractions in the network, simplifying management and facilitating traffic flows.

SDN decouples network devices (switches and routers) as data plane and control planes. The control planes contains network intelligence (controller) to manage signal messages and data forwarding devices. SDN architecture provides a global view and an increased level of network programmability to create flexible capabilities and provide effective QoS provisioning mechanisms. The QoS provision architecture leverages SDN`s class of service capabilities. It enables QoS requests and negotiations between applications and the SDN controller.

However, the solution negotiates small, medium and large traffics with subscribers and service providers; traffic classification is not supported by the the current dynamic traffic management mechanism.

In contrast to the architecture-based approach, recently, machine-learning algorithms have recently gained more recognition due to their high performance in achieving good QoS requirements [41]. Machine learning algorithms such as random Forest, Naïve bays, neural network and decision tree showed better performance in classifiying traffic in data centers for deploying IOT and fog platforms for processing bills and other similar transactions [13] [30]. As the goal of machine learning is to identify sample data and build a learning model, it classifies the testing samples through the constructed classifier.

Many neural network algorithms such as recurrent neural networks of different types are used for classifying network traffic, focusing on feature selection and extraction [30] [40]. Since feature and algorithm selections are very important for improving the classifier performance, SDN-based features such as actions, flow size, open flow protocol, duration and application type are very important in addition to the usual five-tuple parameters such as source address, destination adress, source port, destination port, protocol.

Other authors also tried to mix machine learning and neural network algorithms to solve raffic identification problems. In this regard, Shi Dong and Ruixuan Li [42] proposed a novel application identification method with multiple neural network layer to improve the efficiency and flexibility of application identification. A single application is treated in a single neural network module. Naïve baye algorithm was integrated within a single neural networks module to classify traffic further from a single application.

Bayat and et al[43] identify the traffic flows of applications and services. Traffic classification involves extracting high level features from network packet data and then training with CNN based on packet payload and inter-packet arrival time parameters. CNN was applied to online ML services, offline ML services to achieves users' QoS guarantee and high system utilization. Based on a prediction model a QoS-guided scheduling strategy can be proposed to identify the best placements for traffic.

Mosab and et al [11] also conducted a study on traffic flow management in SDN networks. An elephant detection technique was used to create two classifiers for SDN switches. In their work, most mouse flows in the switches can be filtered based on sketches statistics. Therefore, for mouse flows can not send requests and signaling messages to the controller to minimize the load on controller. Therfore, elephant flow detection becomes more interesting and dynamic when modeled with deep learning algorithms [35]. Deep learning is a branch of machine learning of neural networks [44], which has better learning ability for highly complex tasks than machine learning [36].

Malik and et al [37] used deep learning algorithms including deep neural network in SDN. In their paper, network traffic identification is an essential function for fine-grained traffic management task, although application classification based on application type can not always detect elephant flows. their study did not consider the detection of traffic from future fabricable applications, although the performance of the existing application classifier reaches 96% in terms of accuracy. The classifier was integrated with the controller, which overloads the controller. Instead, the flow classifier module can be integrated in openswiches to share responsibility of the controller. Only elephants flows should be better reported to the controller for further elephant flow clusters. Moreover, the task of optimal route selection and assignment can be done by the controller as usual. Thus, the classification in the OpenSwitches and the clustering in the controller can work together to optimize QoS in SDN. The classifier that improves QoS needs to be more dynamic and use deep learning algorithms. The state of the art is deep learning algorithms such as deep neural network [37], CNN [43], LSTM [45] and Deep Autoencoder, which have been tested for traffic classification.

Manuel Lopez-Martin and et al [46] presented the potential of the CNN algorithm for classification tasks, as we intend to implement it for our elephant flow detection. CNN is a classification algorithm that can initially be used authomatically for representative filtering of traffic. by concatenating mulptiple CNNs, including droupout layer, maxpooling, bach normalization layers, complex features can be extracted automatically. The dropout layer provides regularization (a generalization of results for unseen data) by omitting (setting to zero) a certain percentage of the output from the previous layer. This allows the network to not rely too heavily on a particular input, preventing overfitting and improving generalization. The max- pooling layer selects the maximum value of the traffic value to reduce the number of features and the computational complexity of the network. The result is representative output with less sampling. Batch normalization speeds up training and can improve performance results.

According to Ren-Hung Hwang [45] , LSTM used for classifing traffic from many modern software systems and applications for heterogeneous services in the data centers(clouds). The large growth in the number of these services has led to a critical criterion of QoS, which includes factors such as response time, location and cost. As the value of dynamic QoS attributes vary with time, there is a need of advanced algorithms such as LSTM to accurately forecast future QoS values to identify routes or know a service may be about to fail I n advance. Therefore, LSTM-based neural network is very impotant to forecast future QoS values.

According to Edian F. Franco and et al [47], an autoencoder can be classified into four types depending on the structure of the deep learning layer and the regularization. These are: vanilla autoencoder, denoising autoencoder, sparse autoencoder, and variational autoencoder19]. The one we concerned is the variational AutoEncoder (VAE). VAE is a deep generative model that can simultaneously learn a decoder and an encoder from data. An attractive feature of the VAE is that it estimates an implicit density model for a given dataset via the decoder. While learning a generative model for data, the decoder is the key object of interest. The encoder extracts useful features from dataset and learning a good representation. Learning good data representations before building the models is very important. This is the reason that deep learning, in particular VAE solves the fundamental problems of machine learning algorithms to transfer to new training tasks.

The loss function of the autoencoder compares these predictions with the targets and produces a loss value during compression at the autoencoder bottleneck [19]. The result of the comparison is the loss value, which is a measure of how well the network's predictions match the expectations. The optimizer uses this loss value to update the weights of the network [26].

Convolutional Neural Networks (CNN) with variable autoencoders have shown remarkable classification performance. However, the CNN model is susceptible to noise and redundant information encapsulated in the high-dimensional raw input data, resulting in unstable and unreliable predictions. This problem can be solved by using auto-encoders, which are unsupervised dimensionality reduction techniques that filter out noise and redundant information to produce robust and stable feature representations [48]. The experimental result showed that autoencoder based binary classification enables to score an average precision of 97.49% after 10-fold cross-validation of elphant and mice flows [48]. Therefore, CNN based AE can be one of the promising elephant flows detection algorithm in addition to pure supervised algorithms including DNN,CNN ,LSTM and other deep learning algorithms for sake of QoS optimization.

A Summary of previous works is presented in Table 1. the First group used deep learning(DL) techniques in line of traffic classification(TC), QoS based datasets,real time applications and its state of the art deep learning algorithms. The second group used machine learning(ML).The third taxonomy used architecture-based solutions which employed network architecture for network management optimization such as software defined articture(SDN). The black circle in the last column indicates the focus of the work and the last row concerns our work in this paper.

**Table 1.** Summary of previous works

| Paper | Networking Problem | Approach | Method/classifier technique | Application type | Dataset used | Experimental result | SDN | ML/DL | Elephant | datacenter |
|---|---|---|---|---|---|---|---|---|---|---|
| Aceto, Giuseppe, et al [30]. | Encrypted TC | Deep Learning | SAE, CNN, LSTM, | HTTP traffic | 300k Mobile datasets activity | SAE outperforms in TC | o | • | o | o |
| Alfredo, et al[35]. |  | Deep Learning | 1D-CNN | WeChat | MIRAGE 2019 dataset | global model interpretationr | o | • | o | o |
| Malik and Etal[37] | Intelligence TC | Deep Learning | Deep neural network | HTTP, ,Mail, Multimedia | Moore dataset | Better accuracy, precision, recall, and FScore results. | o | • | o | o |
| hi Dong, Ping Wang, Khushnood Abbas[39] | Deep Learning Aplication | Deep Learning | SC-CNN | Image segmentation | MNIST dataset | deep learning create more powerful optimization methods | o | • | o | o |
| Tao Peng[38] | Abnormal traffic detection | Deep Learning | Kmean, AE,Reinforcement | abnormal traffic | NSL-KDD and AWID datasets | achieved good result in time complexity | o | • | o | o |
| Tao Peng and et al[36] | Tc | Deep Learning | CNN,GAN | FTP ,Gmail, , Skype | USTC-TFC2016 dataset | Provide performance than general ML | o | • | o | • |
| Bovenzi, Giampaolo, et al.[24] | model parallelism TC | Machine Learning | RF ,DT,bayes | IoT ,Fog platform | Anon17 NIMS dataset | reducing training time | o | • | o | • |
| Liu, Chang, et al[40] | Encrypted TC | Machine Learning | RNN-AE | Github, ,gmail,Icloud | Real-world dataset 18 applications | Achieves 99.14% performance | o | • | o | o |
| Ibrahim [66] | online Game | Machine Learning-Mixed | Fixed Java Code | http, FTP Skype | LOL game dataset | produces 91% accuracy | o | • | o | o |
| Shi Dong[20] | Multi-class TC | machine Learning | SVM | http, http, imap, dns | MOORE and NOC | improve classification accuracy | o | • | o | o |
| | | | | | datasets | | | | | |
| Yongfeng Cui, Shi Dong, and Wei Liu[54] | Machine Learning Aplication | machine Lerning | Netflow Extended machne learning | FCBF algorithm | Machine Learning Aplication optimization | Algoriythm called FCBF better perfomance | o | • | o | o |
| Shi Dong and Ruixuan Li [42]. | Traffic Identificatin | Machine Learning | Neural network and Naïve Bayes | TCP And UDP flows | MOORE and NOCSET | achieve 95% identification accuracy. | o | • | o | o |
| Shi Dong[13] | Online Encrypted | Machine Learning | Naïve Bayes | online Skype | Skype-SET | reduces false positives and false negatives | o | • | o | o |
| X. Zhang, D. Zhou S. Dong[19] | Traffic identification | Algorithm Based on Architecture | high identification accuracy | Routing Application | NOC_SET ,CAIDA ,LBNL_SET | High Accuracy | o | • | o | o |
| Alexandre and et al. [2] | QoS | Architecture | Fixed Python program | Distributed Applications | small text, video dataset | low overhead, | • | o | • | • |
| Mosab and et al. [11] | Load balancing | Architecture | Fixed Python program | d/t Applications | sketch based filter elephants | Good running time and performance | • | o | • | • |
| This paper | Elephant Flow Detector | Deep Learning | DNN,CNN ,LSTM,AE | Real time apps | NIMS,SDN datasets | 98.78% accuracy | • | • | • | • |

### 3. Motivation

Accurate traffic classification is the basis for various network activities, including network traffic management and network security auditing [38]. Network traffic classification and analysis has been performed using port-based, DPI, and machine learning techniques. However, in recent years, the rapid increase in the number of Internet users and Internet traffic has led to network congestion. As a result, both the port-based and DPI approaches are becoming inefficient due to the exponential growth of Internet applications that incur high computational costs. The machine learning approach, especially the deep learning approach, has shown the potential to detect traffic anomalies aligned with SDN traffic control capability. Therefore, we are motivated to develop a deep-learning model for software-defined networks that can accurately distinguish elephant flows from mouse flows. SDN and deep-learning technologies are the state-of-the-art traffic management techniques that we use to detect elephant flows for QoS optimization. This dynamic QoS optimization allows network administrators or ISP operators to dynamically predict traffic to prevent resource underutilization and network congestion due to resource overutilization [45] [49]. Given the current massive volume of traffic, ISPs need to predict the application type of a flow through the Internet in order to secure, monitor, and sufficiently allocate the QoS requirements of Internet users in advance [20].

1) Intuitively, there are several reasons why network traffic classification can benefit Internet users, network administrators, and ISP operators.

2) Developing a dynamic model in SDN can minimize or avoid congestion problems, i.e., the Internet is constrained or unavailable due to inflexible traffic handling. This makes the network more flexible and programmable for network operators. It also ensures better QoS performance [34].

Integrating deep learning models into SDN network minimizes human intervention i.e. it increases automation and network administrators or operators can customize the network in terms of topology, configuration and additional module integration in openswiches and controllers. Thus, it opens door for network administrators to manage the network in their context.

The above intuitions motivate us to explore network traffic classification using deep learning models in SDN. The proposed elephant detection model encourages Internet subscribers to negotiate with service providers according to their QoS requirements.

### 4. Materials and Methods

In this section presents dataset preparation and model developmemt methods, and traffic classification model description and model evaluation techniques.

#### a. Dataset and Prepossessing

The training dataset is prepared from exixting VoIP data, video streaming and audio data transmission. The VoIP data comes from HTTP and GTalk applications from Management and Security Group (NIMS) dataset [50],which contains about 303549 traffic flows records. We also used Unicauca-dataset,a network traffic dataset with 15,001 instances and 75 features. To evluate the QoS provision in SDN network, SDN dataset is used and tested by the selected deep learning algorithms.

The three datasets are modified assuming QoS and stored in a CSV file. The datasets are used to improve the QoS requirements of applications with elephant and mouse flow classes [51]. Classification was based on packet size, duration, flow size, byte count, and application type as heuristics. We added a class parameter that included elephant flow (1) and mouse flow (0) classes as the last column. Specific heuristics we used in categorizing mouse and elephant flows are duration, packet size or flow size, and number of bytes in a flow. Mouse flows take at least 10 seconds on average [52]. Each

Short flow requires less than approximately 15 packets [53] and each packet contains 500 bytes [15]. The data preprocessing is accomplished based on parameterization of real-time and non-real-time applications. During and after data processing, packet sampling issue has been widely done [39]. Accordingly, We used stratified 10 fold cross validation [12] to evaluate the performance of the models with new unseen traffic predictions. 10-fold cross validation method is used commonly for precision with the premise that a traffic dataset is divided into 10 parts, 9 of which constitute the training data with 1 representing the test data [13].

## A. Feature selection and extraction

It is important to take account the impact of traffic features and application categories for the development of classification models. packets payload information becomes big obstacle to identify QoS based traffic flows. Instead, we have to see network traffic in line of flows by coralating nany features and application catagories. When feature correlation is 0, it represents that two variables are independent, if feature correlation is 1, it shows two variables have a strictly functional relationship [54]. Combining two or more feature characteristic such as packet size, traffic duration, types of applications can yield better flow classification accuracy.

Network traffic is represented by flow-based features. We used the feature prepared for NIMS dataset as an example [50] and added an additional column for elephant and mouse flow classes. We used the packet size threshold used by cisco systems in data centers to determine dynamic thresholds[53]. Cisco referred to as an elephant flow if the flow contains more than 15 packet sizes i.e. short flow is less than 15 packets. We also consider byte size of to refer to the flow as elephant or mice. Packet sizes are typically greater or equal to 500 bytes per packets[53]. Mouse flows have a size of 10KB in open switches of datacenters [15] [52] and an average duration of 10s [52].

As can be seen in Table 2, the last row describes elephant and mice categories. Network flow is described by a set of statistical features which can be calculated from one or more packets of flows and compute feature values [50].

Table 2: Attributes of NIMS dataset

| Attribute | Duration and size of the flow |
|---|---|
| Min forward inter-arrival time | Min backward inter-arrival time |
| Std deviation of forward interarrival times | Std deviation of backward interarrival times |
| Mean forward inter-arrival time | Mean backward inter-arrival time |
| Max forward inter-arrival time | Max backward inter-arrival time |
| Min forward packet length | Min backward packet length |
| Max forward packet length | Max backward packet length |
| Std deviation of forward packet Length | Std deviation of backward packet Length |
| Mean backward packet length | Mean forward packet length |
| Duration | The time taken for arrival of packets |
| Protocol | Application layer protocol(HTTP and GTalk) |
| total_fpackets | Total Number of Bytes in forward direction |
| total_fvolume | Total voume of Bytes in forward direction |
| total_bpackets | Total Number of Bytes in backward direction |
| total_bvolume | Total Number of Bytes in forward / Total voume of Bytes backward |
| Class | Elephant and mice labels |

### b. Deep Learning Techniques

Deep learning techniques became popular due to The explosive growth and availability of data(big data) and the increase of high-performance computing hardware such as graphics processing unit(GPU) to train large amounts of data [55]. It takes longer training time to yield higher accuracy due to its ability to process a large number of features [56].

Deep learning algorithm passes the data through several training batches and layers to yield complex correlation(models) between features [56].

Deep learning models have been recently studied for network traffic classification learnings. At present, the deep learning techniques includes deep neural network(DNN), convolution neural network(CNN),long short term memory(LSTM), deep autoencoder, deep Boltzmann machine, generative adversarial networks and so on [21].

### c. Proposed Traffic Classification models Description

Network traffic classification is a key component for network management and QoS management. Therefore, employing deep learning methods can distinguish network traffics from distributed and multimedia applications [57]. The proposed traffic classification models enable us to obtain better classification results and reduce the classification time through overall optimization without excessive manual intervention,specially in deep autoencoder model [58].

The proposed traffic classification models are developed with deep learning algorithms. Deep learning algorithms including DNN, CNN, LSTM and Autoencoder, use loss function and optimizer components to build and evaluate the elephant detection model. The elephant detection model training continues until the final classifier is built as per epoch(upto 50), batch size settings(128). The final classifier is obtained after many updates of weights.

The selected algorithms usually yields better classifcation performance than other general machine learning algorithms [50]. Therfore, we used these deep learning algorithms to categorize traffic flows into elephant and mouse flows based on tangible attributes including flow size, flow size, total packet size, protocol type,application type and flow duration as heuritics information for QoS provision. Elephant flows are flows that take long time and have large packet size Whereas mouse flows are those with relatively low sizes transmitting for a short duration [15]. Deep learning categorizes not only elephant flows and mmouse flows but also it helps to further cluster elephant flows for more manageable traffic flows. Proper deployment of a traffic load analysis provides valuable insights, including how busy a link is, the average delays, and the average packet size for wise use of path resources. Thus, deep learning based traffic classification model have advantages in minimizing time complexity and achieved good results on top of QoS datasets [38]. The selected deep learning algorithms for building the proposed traffic classification model are discussed . In paticular, the deep Autoencoder is more described and demonitrated diagramatoically since it is one of the cost effective method due to its authomatic training capability in unsupervised manner with out human intervention.

#### 1) DNN

DNN is typically feed forward networks type of multi-layer perceptron (MLP) in which data flows from the input layer to the output layer in one Forward direction without going backward [25]. Training a neural network comprises sequence of layers, which are combined into a network having the input data (traffic flow) and corresponding expected targets (elephant or mice). Some formulated neural networks models used to identify real-time application layer protocol and the work yields lower time and space complexity [42]. These DNN components create chains and map the input data to predictions. The Layers layer 1, layer 2 and layer n are fundamental data structure which is the building blocks of deep learning model formulation. Models are networks of layers that can be formula or algorithm [26]. Layers can be input, hidden (dense) and output. Each layer can have different number of neurons and it can be calculated using parameters, input (i).Wight (w) and bias (b), and output(y) as it is formulated in equation 1.

$$y = iw + b \qquad (1)$$

Activation function enables the DL model to learn complex patterns. The most frequently used activation functions is ReLU which we use it for our traffic classification model [59].

The gap between actual output(y) and expected output(y`) is recorded as loss. The loss function compares the prediction values to the target values to produce a loss value. The model is the classifier (elephant flow detector) which predicts the unseen flow to their target category as per the assignment of class by data expert manually (true target).

The loss value shows us how well the network's predictions match what was expected. If the loss value is high, the optimizer updates the weights to minimize the loss value. The loss function sets feedback signal for learning purpose. We measure not only the loss incurred during training, but also the performance of the model in terms of accuracy.

The optimizer is used to determine how learning proceeds by letting weight update.

### 2) CNN

Feed forward neural network enables end-to-end training from input to out layer by exploiting, exploiting existing deep learning technology [60]. CNN requires an input filtering function and data pre-processing mechanism by defining local correlation between network neurons of neighboring layers to deliver abstract representations of the input features [26]. It is one of the deep learning algorithms which has 1D, 2D and 3D maxpooling filters to reduce the network scale and further reduces the computation load on the pooling filter. We used 1D traffic data and local features are combined to form the global features; then, the pooling filter is used to remove the unnecessary information to obtain abstract data of reduced size [23].

So we can drive that CNN is a special MLP which we employee for elephant flow detection. A normal CNN model consists of different types of layers which allow the model to learn and extract features relevant to classes [61].

**Convolution layer:** This is the layer where n number of filters is applied to extract features based on the given size of the kernel.

**Batch Normalization:** Batch normalization is used to normalize the output of one convolution going as an input to another convolution. This results in efficient training and helps in reducing over fitting [62].

**Max Pooling:** Max pooling layer is used to reduce the dimensionality of the feature map by selecting the maximum value of a particular region based on the kernel size.

**Dropout layer:** This layer is used to reduce overfit by dropping the specified percentage of features from the model. If dropout (noise) is being applied at training time, random changes in the model can be happened at training time and are working to prevent overfit during training. If this is not being applied at validation time, then validation accuracy could be higher than training accuracy [62].

**Fully connected layer:** This is the final dense layer that is mostly used at the end of the network for classification. Unlike pooling and convolution, it has a global operation capability. It takes input from feature extraction stages, globally analyses the output of all the preceding layers and classify the traffic data as elephant(1) or mice(0) [25].

### 3) LSTM

Recurrent neural network (RNN) provides the capability of "short-term memory" which allowed the use of the previous information at a certain point only to be used for the present training task. LSTM is an extension of RNN which provides the capability of "long term memory" where a list of all of the previous information is available for training the current neural neuron [27]. LSTM's main components are the memory cells and the input, forgets, and output gates. These components allow the LSTM network to have connections from previous time steps and layers, where every output is influenced by the input as well as the historical inputs [28].

LSTM uses feedback loops which lets weight update for correct class assigmemnt during model training [27]. So, in this work, we also employee LSTM for traffic classification task. Elephant flows can be detected using multiple LSTM layers, where each layer comprises many LSTM units, and each unit comprises input, forget, and output gates. To prevent overfit problem, we used weight regularization (Adam), dropout, optimal epoch and appropriate batch normalization techniques [24].

### 4) Autoencoder(AE)

AE is categorized under unsupervised learning algorithm. It is a type of neural network that does not require the human labeling of data.

It reconstructs the input to an output of fewer dimensions [63]. The AE is trained to reconstruct its input through a bottleneck layer with fewer dimensions than the data space. The input (training data) and network maps together have different layers including input, encoders, bottleneck, decoders and output. The AE first encodes the input to a hidden representation (code) of lower dimensions and then decodes it back into a reconstruction.

Even though AE is a clustering algorithm, it can be used for traffic classification by learning dynamic threshold value. The threshold value is obtained dynamically from model learning process. The threshold value identification is the first task of AE during classifier formulation. So, class labels are not necessary for elephant and mice classification, rather, class labels are dropped from the datasets unlike training way of DNN,CNN and LSTM. Only threshold criterion is used to detect elephants flows to optimize QoS provision [64] [65].

Elephnat flow detection is accomplished if the traffic size and duration is greater than the treshhold value. For example, if the threshold value is greater than 0.5, the traffic flow will be assigned in the elephant flow category unless otherwise it is a mice flow.

As AE is depicted in figure 1, it takes network traffic data then filters the representative neurons from each instances then,autoencorder continues compressing the filtered data to find threshold value.

Traffic inputs (T) has are given to the autoencoder network then it encodes to coded form with minimal features at the bottleneck layer. The coded feature is decoded to yield the output layer (O). Threshold value is calculated during reconstrcition the network. we use the coded value(treshold) for elephant and mice flow classification task. In Figure 1, Unicauca-Dataset is used to design the autencoder structure. Unicauca-Dataset has 87 attributes and we add one additional class column. Totally,there are 88 parameters used as Input starting $T_1$ to $T_{88}$.

**Figure 1**: Autoencoder for network traffic data

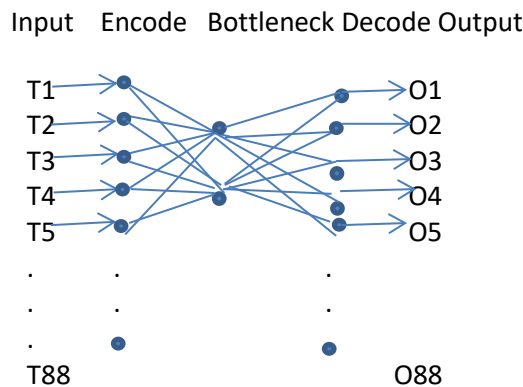

As visualized in Figure 1, we can take an unlabeled dataset and give to autoncoder for learning task. Let us take the original input T as $[T_1, T_2, T_3, T_4, T_5...T_{88}]$ and the output,O as $[O_1, O_2, O_3, O_4, O_5...O_{88}]$, is a reconstruction of the output. Autoencoder can be trained by minimizing the reconstruction error,e(T,O), which measures the differences between our original input and the consequent reconstruction. This construction loss yields threshold value for classification task.

Thus, an autoencoder approximates an identity of the traffic whether the traffic instance is elephant or mice based on threshold value i.e. the hidden layer has two outlets, one for elephant flow and the other is for mice flows [37].

The traffic classification model workflow is also presented in figure 2 having input traffic data(elephant and mice).

Firstly elephant and mice traffic flows are inputed to the workflow. AE Model is then formulated to provide threshold value which is used as the boundry during elephant flow detection i.e the AE model yields the tresh hold value which is a formula to identify elephants in non linear hyperplane manner.

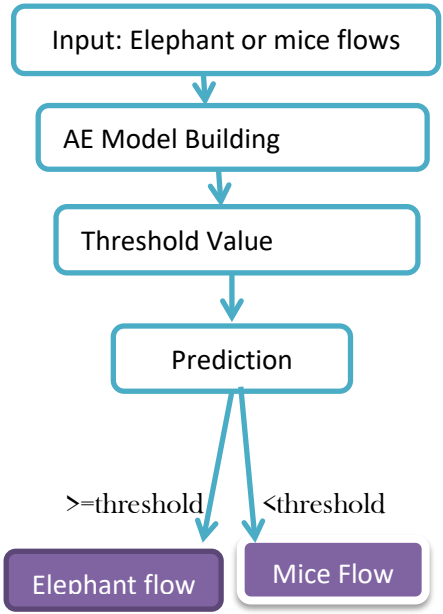

**Figure 2:** Model Prediction using threshold.

Once the formula(hyperplane boundery) is found, traffic prediction is performed based on threshold value. If the traffic input-wight size is greater than this threshold boundery, the flow is detected as elephant as it depicted in Figure 2. Unless otherwise;it is categorized under mice class. So, the network setting does not change because the traffic does not lead to network congestion.

## 5. Experimentation and Evaluation

The Internet traffic measurement and analysis minimizes or avoids traffic congestion challenges [66], but prior to data transmission, the performance of the classifier must be verified with various metrics. We present the results of the four deep learning models on three dataset that we have adapted.. We also compare the performances of the models. To design a robust evaluation framework, we run every model over 20 to 50 epochs. The average performance of the each algorithm is considered as the promising potential of the elephant detection model. We use accuracy and loss metrics for model evaluation. We also checked the effects of overfitting and underfiting of models by conducting experiments on epoch repetitions and training data batch sizes.

### A. Dataset

The goal of this study is to develop a model that is best suited to detect elephant flows. To achieve this goal, we selected an inelastic and an elastic application-based dataset. Thus, we used features from the NIMS dataset and took a non-real-time HTTP application as an example and Gtalk as a real-time application to formulate the model [50]. We then changed the class column as elephant (1) and mouse flow (0). We also used two other SDN-based datasets to test the performance of the proposed model in openswiches. The datasets were divided into training, validation, and testing sets with 10-fold stratified cross-validation.

The training parameters, epochs, batch size, learning rate, and optimizers used in this model are listed in Table 3.

**Table 3:** Training Parameters

| Model | Epoch | Batch | Rate | Optimizer |
|-------|-------|-------|------|-----------|
| DNN   | 20/50 | 128/512 | 0.01 | Adam |
| CNN   |       |       |      |      |
| SAE   |       |       |      |      |
| LSTM  |       |       |      |      |

### B. Experimental setups and tools used

We installed Anaconda version 3 on an Intel(R) Core(TM) i7-4500U CPU @ 1.80GHz 2.40 GHz laptop computer. We also installed components to help us perform the deep learning experiments. Some of the components are Jumpy, Pandas, and Matplotlib to process a variety of data and graph the experimental results. To control traffic and support QoS, we group incoming traffic into elephant and mouse flows. The time required to complete one round of execution of each model is recorded in seconds. The models are built using the Python version 3.9 programming language and the Tensor Flow 2.3.0 and Keras 2.4.5 frameworks.

In Deep Learning, the input data is trained by automatically learning structured feature representations using the Keras framework [37]. Keras is a powerful and easy-to-use Python library for developing and evaluating deep learning models based on Tensorflow. Tenssorflow enables the definition and training of network models as a multidimensional array or list [26] [67]. Keras and Tensorflow modules were installed on Anaconda to obtain deep learning-based autoencoder libraries. We implement DNN, CNN, LSTM, and autoencoder algorithms on SDN datasets. We run CNN and LSTM deep learning codes on Googlecolab to obtain fast computing performance.

### C. Model Evaluation

The proposed model is evaluated using 10-fold stratified cross-validation on a test set. A classification result has four cases: true positive (TP), false positive (FP), true negative (TN), and false negative (FN). For our purpose, we used the same input form, the same training set, the same learning rate, and the same optimizer.

For cross-validation, we stratified the SDN dataset into 10-fold. The SDN dataset consists of a large number of instances. The dataset is automatically split into a training set and a test set. The training and test datasets are partitioned in a stratified manner, starting with fold1 and ending with fold 10, as shown in Table 4.

**Table 4:** Stratified 10-Fold on SDN Dataset for Cross Validation

```
Fold:1, Train set: 93455, Test set:10384
Fold:2, Train set: 93455, Test set:10384
Fold:3, Train set: 93455, Test set:10384
Fold:4, Train set: 93455, Test set:10384
Fold:5, Train set: 93455, Test set:10384
Fold:6, Train set: 93455, Test set:10384
Fold:7, Train set: 93455, Test set:10384
Fold:8, Train set: 93455, Test set:10384
Fold:9, Train set: 93455, Test set:10384
Fold:10, Train set: 93456, Test set:10383
```

To measure the performance, the metrics used is Accuracy which is defined as follows.
Accuracy = TP + TN/ ( TP + FP + TN + FN).

### D) Discussion of Experimental Results

In this experimentation, we use DNN, CNN, LSTM and autoencoder algorithms for our traffic classification. The Experiment results are presented for each research deep learning algorithms used are discussed. We used experimental results of the training history to measure performance of the models in terms of accuracy and loss for each employed algorithms.

Experiment I: to what extent do DNN, CNN, LSTM and autoencoder neural network algorithms detect elephant flows from mouse flows for good QoS?

### 1) Experimenting Elephant flow detection using DNN

Network`s elephant detection model classifies elephants and mouse categories to the target outputs. When we see the performance of the model in terms of accuracy using Adam optimization, training accuracy increases and training accuracy reaches to 99.99 % under epoch 50 on NIMS dataset. The validation accuracy reaches 100% under the same epoch with in 3s and 5ms.

The model training history of the model and performance accuracy of DNN algorithm are shown in Figure 3(a). It Shows the training accuracy and validation accuracy of the model. The final traffic classifier model is then verified with the best (i.e., highest) validation accuracy [68].

Figure 3(b) shows the training loss and validation loss of the model. The final model is the checkpoint with the lowest validation loss, 0.0037 which is closer to zero [68].

### (a) Training accuracy and validation accuracy

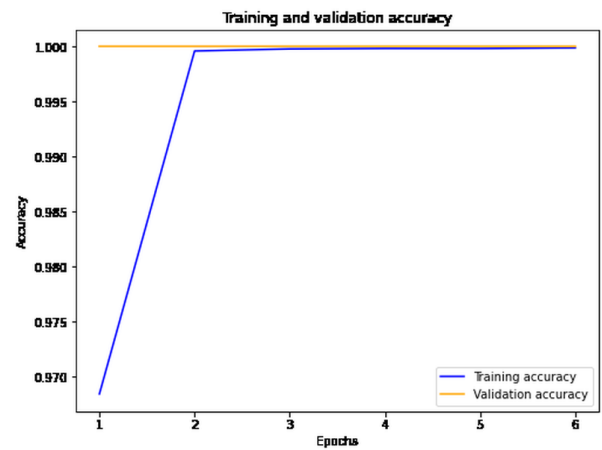

(b) Training loss and validation loss

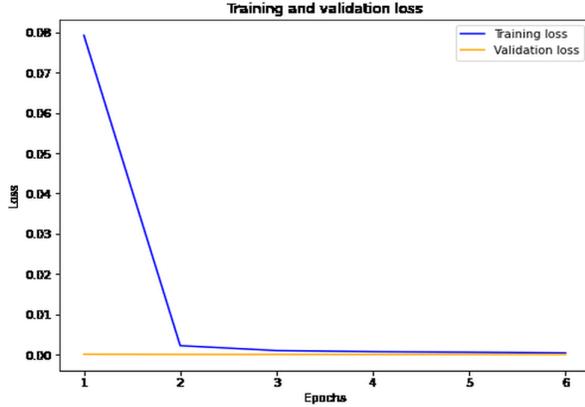

**Figure 3**: DNN Performance Result Using NIMS dataset

The training loss diminishes starting from 0.0793 to 0.0022 as it is seen in figure (b) above. The test validation of the model using 10 fold stratified cross validation is presented in the confusion matric from figure 4.

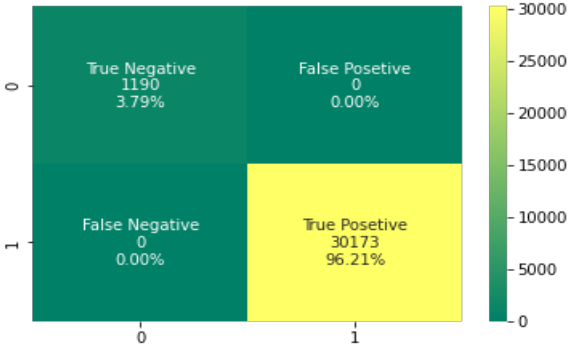

**Figure 4**: Confusion Matrix on NIMS dataset

False positive and false negative errors are reduced to zero by the judicious use of heuristics that take into account the duration of data flow, the size of data flow, the type of application ,and the size of the packets.We also verified the performance of traffic flow classifier with a labeled application dataset, Unicauca- Dataset. The total dataset has 15, 001 instances and 87 parameters. We modified the dataset considering the QoS requirements of applications and stored in a CSV file. We add the 88th clumn as class column, that holds elephant flow (1) and mouse flow (0) classes. The duration of the mouse flow requires at least 10 seconds. Each flow has at least 15 packets and each packet contains 500bytes as it is stated in our methods. Having this heuristic information, we create a learning model that can detect elephant flows. The model performs 97.36% training accuracy with in 7ms and 97.24% testing accuracy using deep neural network on unicauca- dataset [69] as it is seen in figure 5(a). The validation accuracy was constant while the training accuracy increases radically starting 70.03% to 96.52%. The loss that occurred during training on Unicauca-Dataset showed radical reduction starting from 0.6443 to 0.1199 as shown in figure 5(b). As training loss is presented, the training loss scored approaches to zero as the final model is the checkpoint with the lowest validation loss.

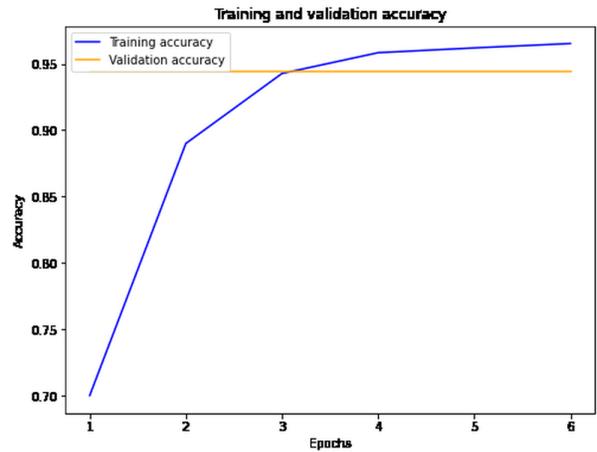

(a) Training accuracy and validation accuracy

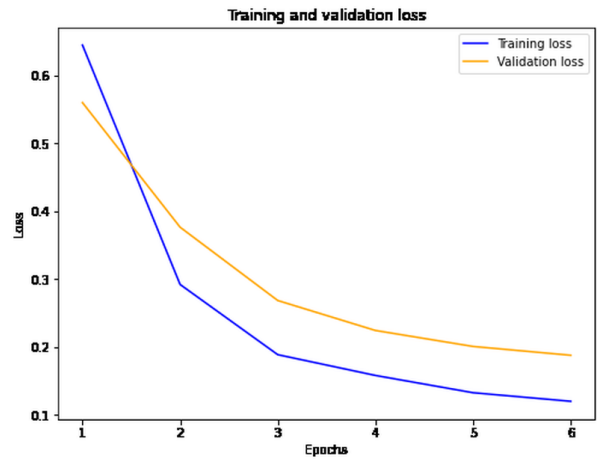

(b) Training loss and validation loss

**Figure 5**: DNN loss on Unicauca- Dataset [69]

The validation of the model is demonstrated by confusion matrix in figure 6.

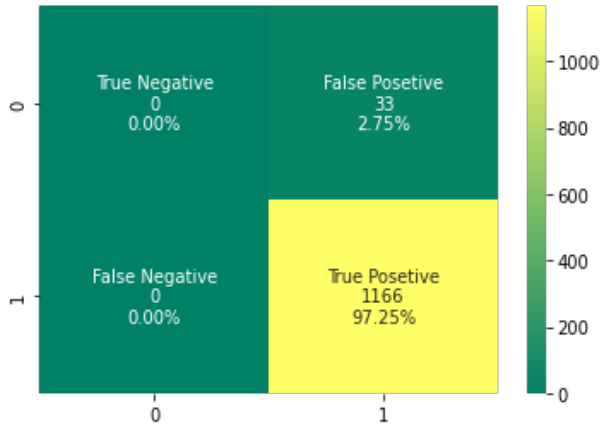

Figure 6: Confusion matrix on unicauca- dataset

The main objective of this work is traffic classification on SDN dataset using state of the art DNN, CNN, LSTM and Autoencoder algorithms. The dataset was generated from RYU controller. It has 104346 instances with 23 features. We modified the last column and substituted by elephant (1) and mice (0) classes after identify duration, packet size, byte size, flow size and application protocols parameter as heuristics.

The DNN experimental result obtained within 50 epochs is 99.97% training accuracy and validation accuracy is 100 % in 1s and 5ms as it is seen in figure (a).

We checked the performance of the classifier by performing 10-flod stratified cross validation with testset during evaluation. We achieved a test accuracy of 100%, which is promising model performance [70]. Interpreted to mean that test accuracy should not be higher than training accuracy. when the model is not overfitted, the training model classifies new samples as the training model is optimized for the testing samples latter [62].

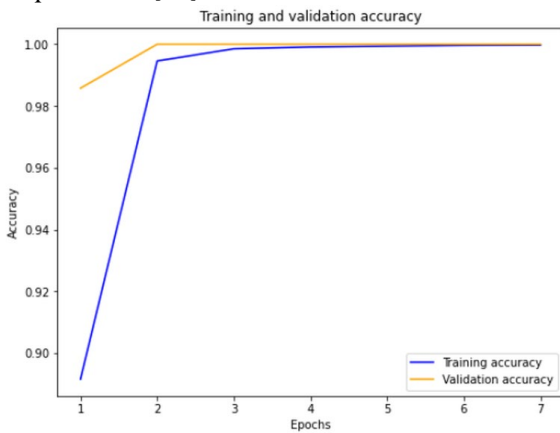

(a) training accuracy and validation accuracy

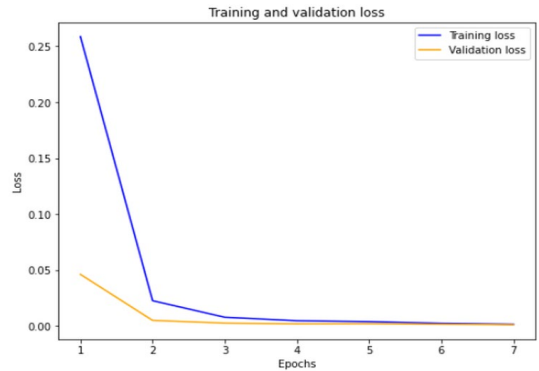

(b) training loss and validation loss

Figure 7: DNN Loss on SDN Dataset [12].

The training loss decreases from 0.2587 to 0.0014 which is generally close to zeroas shown in figure7(b). 0.0014 noise is a normal occurrence, as swe do not expect 100% perfection from machine, as humans have natural limitation to perform 100% autonomously. Humans detect during the day and night with accuracy of 80% average. The overall accuracy is usually expressed as a percent, with 100% accuracy is being a perfect model, achieving 100% recognition is very difficultg task for machines [71].

Let us explain our model performance using confusion matrix. The prediction classes of the models are predicted values and actual values along with the total number of predictions [72]. Predicted values are those values, which are predicted by the model, and actual values are the true values for the given observations labeled by us as shown in figure 8.

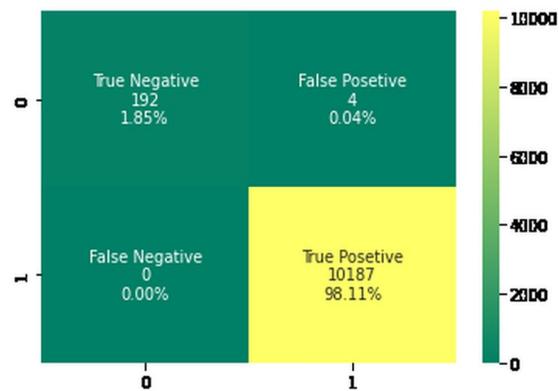

Figure 8: 2×2 Confusion Matrix SDN Dataset using DNN.

**True Negative:** 192 records were assigned to mouse streams as we annotated them as mouse flow.

**True Positive:** we predicted elephant flows were positive (1) and 10187 records are actually truly mapped to elephant flows in datacenters.

**False Negative:** The model predicted elephant and mouse flow without any error.

**False Positive:** The model has predicted 4 mouse records as elephants mistakenly. Thus, the error is only 0.04%.

### 1.1) Discussion on performance result of DNN algorithm

Model accuracy and explanations are relevant for many practical applications of deep neural networks, such as our traffic classification task. Accordingly, we found elephant classification model to be 97.36%, 99.99% and 99.97% on NIMS, unicauca and SDN datasets respectively. When we calculate the training accuracy, the average performance of DNN is 99.12% for elephant detection model. The model performance implies that our elephant detecting model can lead to better generalization performance under DNN [73]. DNNs optimizes classification tasks with respect to training data in a heterogeneous manner as we tested on heterogeneous traffic including datasets generated from legacy and SDN networks. With sufficient parametric flexibility, these types of models can fit generalizable features and memorize non-generalizable features concurrently during training [74]. The built model meets the general standard of deep learning models i.e. the validation and test accuracies are greater than 96% and all validation and test accuracies were less than the training accuracy [62].

we can infer from the graphs figure 3(b), figure 5(b) and figure 7(b), both training and validation loss are decreasing further,which is generally near to near zero. Little noise is a normal occurrence since we do not expect 100% perfection from machine as human being cannot perform 100% autonomously [71].

### 2) Experimenting Elephant flow detection using autoencoder

The goal of the autoencoder is to find optimal model parameters for the minimization of a loss function. The msle loss function on training samples finds the maximum loss value [75]. For this work, mean squared error loss function (msle) was used with Adam optimizer and the loss approaches to zero as it is seen in figure 9.

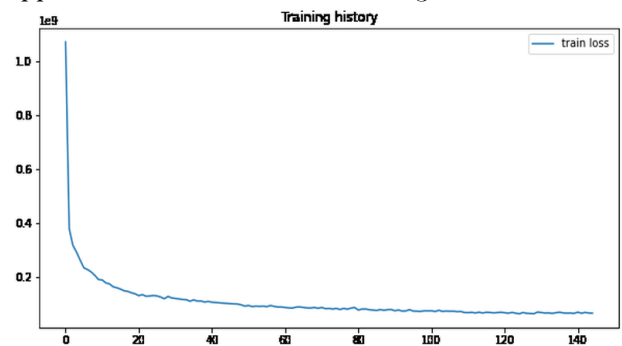

**Figure 9**: Loss visualization Autoencoder

Distribution of the deep autoencoder model, reconstruction loss distribution visualization and the loss decreases uniformly and approaches to zero over training history. In particular,the loss decreases from 1 to 0.13. Based on the train loss, many reconstruction loss values are calculated [74]. The best reconstruction loss helps to find the optimal threshold. Accordingly, we detect elephants by reconstructing the input traffic. The threshold value for elephant flow detection is 0.1555 with model validation accuracy of 97.95% as shown in table 5. If the reconstruction loss for a sample is greater than this threshold value then we can infer that the model is seeing a pattern that it isn't familiar with mouse flows. The test accuracy score obtained is 96.58% accuracy.

**Table 5**: Best validation accuracy and threshold

```
------------------------------------
Best validation accuracy: 0.9795 for threshold: 0.1555
Perentile:90 Threshold: 0.1005 Validation Accuracy: 0.9521
Perentile:91 Threshold: 0.1028 Validation Accuracy: 0.9543
Perentile:92 Threshold: 0.1083 Validation Accuracy: 0.9658
Perentile:93 Threshold: 0.1113 Validation Accuracy: 0.9658
Perentile:94 Threshold: 0.1173 Validation Accuracy: 0.9703
Perentile:95 Threshold: 0.1245 Validation Accuracy: 0.9726
Perentile:96 Threshold: 0.1383 Validation Accuracy: 0.9772
Perentile:97 Threshold: 0.1555 Validation Accuracy: 0.9795
Perentile:98 Threshold: 0.192 Validation Accuracy: 0.9795
Perentile:99 Threshold: 0.2614 Validation Accuracy: 0.8174
----------------------------------------------------------
Best validation accuracy: 0.9795 for threshold: 0.1555
```

### 1.2) Discussion on performance result of Autoencoder

The loss function quantifies how well, or how bad the given predictor when it classifies the input data points in the dataset. The smaller the loss, the better the classifier is doing at modeling the relationship between the input data and output class labels [70]. Loss is a cumulative noise per epoch. At the beginning of each epoch, loss as 9.5. For each time,loss calculation is added to the loss metric [70]. What is seen over time in the plotted results is the total loss of training and validation loss is decreasing generally. This decreasing of loss means the weights of the network is getting more accurate. Usually, reading such a low loss of near zero indicates that the potential of the model to detect elephant flows.

It is implicitly expected that the classification accuracy is inversely proportional to the average loss value i.e. the training loss and validation loss both decrease and stabilize at a specific point a round epoch 20.

Usually, the validation loss is greater than the training loss. This may indicate that the model is under fitting for some extent [62]. Even though the results shows small loss, result indicates that further training is needed to reduce the loss incurred during training for more performance improvement. Alternatively, we can also increase the training data either by obtaining more samples or augmenting the data [76].

3) Experimenting Elephant flow detection using CNN and LSTM algorithms.

We compared the performance of DNN with the state of the art CNN and LSTM algorithms. We ran CNN on SDN dataset and its performance result was 98.17% accuracy and 98.13% validation accuracy. Loss of 1.83% is occurred during training. The total training takes 8 seconds to build the elephant detection model. We also develop the elephant detection model using LSTM. The training performance result scores 98.78% training accuracy and 97.55% validation accuracy respectively, as shown in figure 10 (a). The training takes 57seconds in average with minimum (3.2%) loss,as shown in Figure 10 (b).

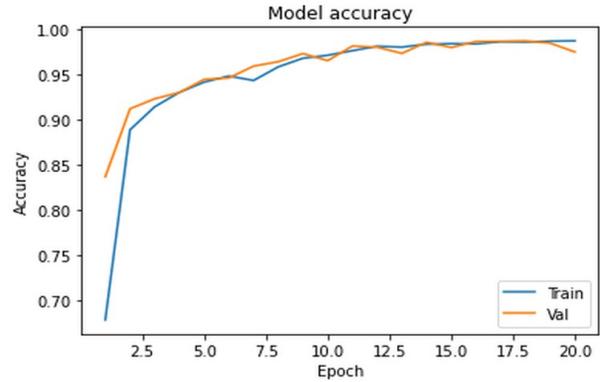

(a) Training accuracy and validation accuracy

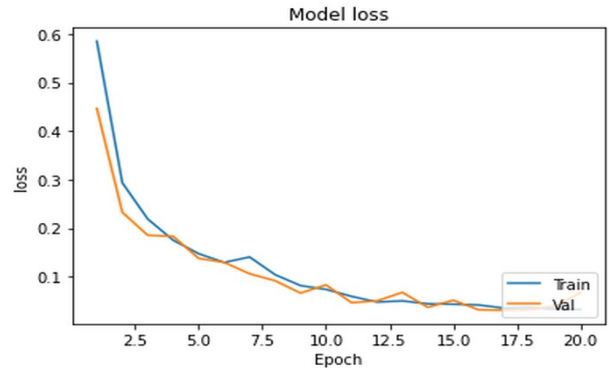

(b) Training loss and validation loss

Figure 10: LSTM Model Performance

1.3) Discussion on performance result of LSTM

A slightly higher increase of validation accuracy over training accuracy up to epoch 11 is due to the over fitting problem and stabilizes the traiing to the normal state after 11 epoches [62]. The LSTM model shows fairly increasing accuracy after 11 epochs. Thereby, the loss value drops from 7.84 % to 0.56% which is near to zero. A low loss of near zero indicates that the elephant and mouse classifier has the of dthe model to categorize the traffic flow to the target traffic class [77].

At the beginning of training, over fitting of the model on SDN dataset was happened. Figure 10 (a) shows that after a certain epoch, the validation accuracy increases while the training accuracy decreases. After epoch 5, the training accuracy becomes higher than the validation accuracy, which is due to the dropout techniques we incorporated.

Here, both the training accuracy and validation accuracy were balanced, which is advisable when building the best models.Figure 10(b) shows the gap between training loss and validation loss over time. Both the training loss and validation loss are close to zero. The scenario ensures that the training was performed in a normal situation that is free from overfitting and underfitting problems. Therefore, LSTM shows good fitting, validation accuracy and training accuracy are high, with the former slightly lower than the latter[62]. LSTM not only shows goodfit, but also achieves the highest accuracy (98.78%) among the other deep learning algorithms used in this work.

Experiment II: Comparison of DL techniques

Deep learning algorithms were compared in terms of their relevant complexity including performance in accuracy, model runtime, number of trainable parameters and loss incurred.

Table 6: Comparison of deep learning Algorith

| Algorithm | Performance In Accuracy | Model run time | Loss | Dataset used | #parameters |
|---|---|---|---|---|---|
| DNN | 99.12 % in average | 4s | 0.041 | NIMS, Unisuaca,SDN | NIMS-24, SDN-21, Unisuaca-75 |
| CNN | 98.49% | 5s 3ms | 0.015 | SDN | 21 |
| LSTM | 98.78% | 57s 60ms | 0.03 | SDN | 21 |
| Deep-Autoencoder | 97.95% | 59s | 0.13 | SDN | 21 |

The optimal performance accuracy of the models is represented in various indicator parameters. 99.12% an average accuracy is the highest performance obtained with DNN. Although this is the highest performance, LSTM, CNN, and Autoencoder also provide promising models with accuracy greater than 96%.

In addition to accuracy, we also compared the time taken to produce the models, as shown in Table 6. From the comparison results, our models generally run in less than 60 seconds. In particular, DNN has the lowest runtime of all the algorithms used, at 4 seconds. Deep autoencoder takes 59 seconds, which is a relatively slow performance due to the additional compression and decompression tasks [72] [78].

The comparison result on three QoS-based datasets, NIMIS, nicauca and SDN, is also presented with the number of parameters being ,24, 75 and 21, respectively. Thus, the dataset preparation and parameterizations were important steps in the elephant flow detection process. Duration, flow size, packet size and application type were used as heuristics parameters. Since the focus of this work is on SDN,the SDN dataset with DNN, CNN, LSTM and Deep Autoencoder was used.

Experiment III: What optimal epochs will be required for producing of the flow classification model?

As we try to find the optimal or correct number of epochs to train a neural network model, we experiment with different epoch numbers and batch sizes to check how accuracy is affected. It is also used to check whether overfitting and undermining of the training occurs when classifying the network traffic. First, we perform the training on the NIMS dataset with 2 hidden layers, an input layer and an output layer, by initializing the epochs to 50.

Epoch is a kind of hyperparameter that plays an essential role in the training process of a model and helps to decide whether the data is overtrained or not [67]. Thus, we find that epochs play an important role in obtaining good accuracy for training the neural network model only on the traffic training dataset [70].

Neural networks are able to learn by changing the distribution of weights. It is possible to approximate a function that is representative of the patterns in the input. The key idea is to stimulate the black box with new stimuli (data) until a sufficiently well-structured representation is obtained [79]. Therefore, the dataset is tested in different epochal iterations. Accordingly, we test different epoch intervals to find the optimal epoch for the dataset NIMS using CNN-based AE, as shown in Table 7.

Table 7: Model performance using different epochs.

| Batch size | Epoch | Accuracy NIMS dataset[30] |
|---|---|---|
| 512 | 5 | 96.08% |
| 512 | 10 | 97.01% |
| 512 | 20 | 98.70% |
| 512 | 50 | 98.74% |
| 512 | 100 | 98.70% |
| 512 | 1000 | 98.06% |

We choose CNN based Autoencoder to assess the effect of epoch on the models having the behavior of mixed algorithms used to construct models in the work: CNN and AE. The result of this mixed algorithm. We can see the accuracy of the models for the six different epoch values. It is shown that the classification result is trained with 5, 10, 20, 50, 100, and 1000 epochs, respectively, with constant stack size. The network computes the errors for both the training and validation sets. We stop training when the validation error reaches the minimum.

### Discussion

As we tried to find the optimal epochs to train a deep learning, an experiment was conducted with various numbers of epochs to check how the accuracy affects the model performance as well as if there is any over fitting happening in the training process.

As the results shown in Table 7, as the epoch value increases, the accuracy of the system is improved starting epoch 5 to epoch 50.

We can conclude that the accuracy of the model is promising for detecting elephant flows with the same stack size and number of classes, so most models have good accuracy for each epoch [77]. However, performance begins to decline after epoch 50. This is due to the fact that up to epoch 50, most records are classified into their category based on the optimal updated weights. Thus, the classification saturates when epoch is selected until about epoch 50, then the classification accuracy decreases because the traffic flows are already assigned to their class, i.e., the updated weights are summed up beyond the expected computational result [70]. In other words: If the accuracy of the training data increases, but the accuracy of the validation data remains the same or even decreases, it means that the model is overfitted, which in turn means that we should stop the training process [79]. Thus, we can conclude that the validation error is increasing. At epoch 1024, the accuracy becomes high (98.22%), but this is an overfitting since the training error is equal to 0 [70]. So we have to stop at epoch 50 for sure. We can stop the training process early at about epoch 50 to get better performance of the model. Since epoch 50 has achieved 98.74% accuracy, this is optimal and promising for integration with SDN controllers or OpenSwiches for traffic management to improve QoS.

Experiment IV: By what extent batch size variation affects traffic flow classification model at constant epoch?

We test the performance of the elephant detection models when the lot size varies. As shown in Table 8, the best model performance is achieved at a batch size of 512, while the epoch is constant at 50 when using CNN-based AE.

Table 8: Different batch size on model performance

| Batch size | Epoch | Accuracy |
|---|---|---|
| 32 | 50 | 97.01 |
| 64 | 50 | 94.30 |
| 128 | 50 | 97.60 |
| 512 | 50 | 98.74 |
| 1024 | 50 | 98.22 |

The batch size indicates how much of the data set is used for each training step. The training process is a stepwise process where the dataset is divided into batches [77]. In our case, we tested our dataset by dividing it into 32, 64, 512, and 1024 batches. The stack size of 512 achieves the best performance, 98.74% accuracy, as seen in Table 5.6 above. We tested the variation of stack size to understand its impact on the traffic model construction. We tested the stack size starting from 32, 64, 128, 512, 1024 under the constant optimal number of 50 epochs. The stack size with 50 epochs gives an accuracy of 98.74%. The best results in traffic classification accuracy are obtained with stack size of 512 and 1024 examples. The larger the stack size, the higher the traffic classification accuracy [80]. It can be concluded that a model with a stack size of 512 examples and an epoch of 50 examples is promising for a small traffic dataset and shows the potential of the QoS provisioning mechanism.

## 6. Conclusion and Future Work

### A. Conclusion

Deep learning techniques have become one of the most interesting and practical topics in network engineering. In this paper, we propose a traffic classification model that detects elephant flows and can be integrated with SDN controllers to ensure good QoS. In particular, we used deep neural networks, CNN, LSTM, and autoencoders. To consider a model as the best model, its performance was tested with the training dataset, validation dataset, and test datasets. The flow classification model was found to be the most influential model for classifying elephant and mouse flows in SDN. The average detection rate of elephant fluxes is 98.77%, 98.17% and 98.78% using DNN, CNN and LSTM in the three datasets, respectively. Therefore, we can conclude that the potential and capabilities of deep learning algorithms for elephant flow detection are promising for better QoS in SDN.

Therefore, we can conclude that the potentiality and promising of deep learning algorithms on elephant flow detection for better QoS in SDN.

### B. Research Implication

Deep learning techniques have become one of the most interesting and practical topics in network engineering. In this paper, we propose a traffic classification model that detects elephant flows and can be integrated with SDN controllers to ensure good QoS. In particular, we used deep neural networks, CNN, LSTM, and autoencoders. To consider a model as the best model, its performance was tested with the training dataset, validation dataset, and test datasets. The flow classification model was found to be the most influential model for classifying elephant and mouse flows in SDN. The average detection rate of elephant fluxes is 98.77%, 98.17% and 98.78% using DNN, CNN and LSTM in the three datasets, respectively. Therefore, we can conclude that the potential and capabilities of deep learning algorithms for elephant flow detection are promising for better QoS in SDN.

### C. Future Work

In our future work, we plan to conduct a research on SDN by integrating the best model obtained in this work so that we will review the QoS and QoE improvements between users and network administrators. In the future, we plan to increase the amount of data and extend this research study by using different deep learning methods in an SDN environment. This will provide an opportunity for very accurate, fast and reliable classification. In particular, the elephant flow detection task should be better tested with a generative adversarial network (GAN) since GAN shows good performance in pattern recognition. Generating adversarial deep convolutional networks helps for efective fits and expands traffic dataset to maintain balance between elephant and mice classes of the dataset, which enhances the dataset stability [36]. In addition, the effect of explainable artificial intelligence (EAI) to improve the quality of service in SDN networks is investigated.

### Data Availability

The NIMS QoS dataset is available at the link: https://projects.cs.dal.ca/projectx/Download.html.

The Unicauca_Dataset is available at the link: https://www.kaggle.com/datasets/jsrojas/ip-network-traffic-flows-labeled-with-87-apps
The original SDN_traffic dataset is available at the link:https://github.com/Rachico/deep SDN/blob/main/sdn_traffic_dataset.csv

These dataset is modified to identify elephant and mice flows. The modified dataset is available at gethub:
https://github.com/getahunwassie/Researches-papers

## Conflicts of Interest

No conflicts of interest happened between authors during and after the writing and publishing of the paper
.

## Acknowledgments

Our acknowledgement goes to authors of the dataset [50], [69] and [81] who prepared traffic datasets and make it available openly.

## Reference

[1] Jonathan, et al. Falk, "Dynamic QoS-Aware Traffic Planning for Time-Triggered Flows in the Real-time Data Plane," *IEEE Transactions on Network and Service Management*, 2022.

[2] " Oliveira, Alexandre T. et al.," *"SDN-based architecture for providing QoS to high performance distributed applications."*, 2018.

[3] Hamza Awad Hamza, Sulaiman Mohd Nor, and Ali Ahmed Ibrahim, "Internet traffic classification algorithm based on hybrid classifiers to identify online games traffic," *Jurnal Teknologi 64.3*, 2013.

[4] Bilal Khalida and Wornchanok Chaiyasoonthorna Singha Chaveesuka, "Chaveesuk, Singha, Bilal KhaContinuance intention to use digital payments in mitigating the spread of COVID-19 virus," *International Journal of Data and Network Science*, 2022.

[5] Singha, Phayat Wutthirong, and Wornchanok Chaiyasoonthorn Chaveesuk, ""Cloud computing classroom acceptance model in thailand higher education's institutes A conceptual framework," *Proceedings of the 2018 10th International Conference on Information Mana*, 2018.

[6] Ashkan Gholamhosseinian and Noushin Zare Hajibagher Ahmad Khalifeh, "QOS For Multimedia Applications with Emphasize on Video Conferencing, , hp Halmstad University,School of IDE," *Modern Communication System and Networks*, Feburary 2011.

[7] Chen Wang and Hyong Kim, "Touchdown on the Cloud:The impact of the Super Bowl on Cloud ," vol. 1902.07363v1 [cs.NI] , Feburary 2019.

[8] Feng, et al Mao, "Hierarchical video frame sequence representation with deep convolutional graph network ," *Proceedings of the European Conference on Computer Vision (ECCV) Workshops*, 2018.

[9] Yutong Zhai and at.al, "Joint Routing and Sketch Configuration Membe," *, IEEE IEEE/ACM TRANSACTIONS ON NETWORKING*, vol. VOL. 28,no 5, OCTOBER 2020.

[10] Anshuman Chhabra and Mariam Kiran, ""Classifying Elephant and Mice Flows in High-Speed Scientific Networks"," *,Division of Electronics and Communication Engineering*, September 2017.

[11] Bushra Mohammed, Usman Humayun, Ahmed Abdelaziz, Suleman Khan, M. Akhtar Ali, Muhammad Imran, and Muhammad Nadzir Marsono Hamdan Mosab, ""Flow-aware elephant flow detection for software-defined networks"," *IEEE Access* , 2020.

[12] Fan and R. Liu, ""Investigation of machine learning based network traffic classification"," *International Symposium on Wireless Communication Systems (ISWCS)*, 2017.

[13] Shi Dong, "Online Encrypted Skype Identification Based on an Updating

Mechanism," *arXiv preprint*, 2022.

[14] Mahdi Jafari Siavoshani, Ramin Shirali Hossein Zade and Mohammdsadegh Saberian Mohammad Lotfollahi, *"Deep Packet: A Novel Approach For Encrypted Trafic Classication Using Deep Learning"*, July 2018.

[15] Adibi Sasan, ""Traffic Classification, Packet-, Flow, and Application-based Approaches"," *(IJACSA) International Journal of Advanced Computer Science and Applications*, 2020.

[16] Behzad Akbari, and Nader Mokari Mohammad Mahdi Tajiki, ""Optimal QoS-Aware Network Reconfiguration in Software Defined Cloud Data Centers"," *Computer Networks*, vol. Volume 120, June 2017.

[17] Yotam Harchol, David Hay and Yaron Koral, Anat Bremler-Barr, "Deep Packet Inspection as a Service," *Proceedings of the 10th ACM International on Conference on emerging Networking Experiments and Technologies*, December 2014.

[18] Shi, Dingding Zhou, and Wei Ding. Dong, "Traffic classification model based on integration of multiple classifiers," *Journal of Computational Information Systems*, 2012.

[19] X. Zhang, D. Zhou S. Dong, "Auto Adaptive Identification Algorithm Based on Network Traffic Flow," *International Journal Of Computers Communications & Control,ISSN 1841-9836, 9(6):672-685.*, December 2014.

[20] Shi. Dong, "Multi class SVM algorithm with active learning for network traffic classification," *Expert Systems with Applications*, 2021.

[21] Shi, Ping Wang, and Khushnood Abbas Dong, "A survey on deep learning and its applications," *Computer Science Review 40*, 2021.

[22] Hang Zhu, Xin Jin and Ion Stoica Eric Liang, ""Neural Packet Classification"," Feburary 2019.

[23] Imtiaz Ahmad and Ebrahim Alrashed Dalal Aloraifan, ""Deep learning based network traffic matrix prediction"," *International Journal of Intelligent Networks 2*, 2021.

[24] S. S., Jabbar, M. A., & Tiwari, S Raoof, "Foundations of deep learning and its applications to health informatics," *In Deep Learning in Biomedical and Health Informatics*, 2021.

[25] Asifullah Khan and et al, "A Survey of the Recent Architectures of Deep Convolutional Neural Networks," *Published in Artificial Intelligence Review*, 2020.

[26] Chollet rançois, ""Deep Learning with Python"," 2018.

[27] Yuanyuan Wei and et al, "LSTM-Autoencoder based Anomaly Detection for Indoor Air Quality Time Series Data," 2022.

[28] Tareq Tayeh and et al, "An Attention-Based ConvLSTM Autoencoder with Dynamic Thresholding for Unsupervised Anomaly Detection in Multivariate Time Series," *ECE Department, Western University, London, ON N6A 3K7, Canada; saburakh@uwo.ca (S.A.);*, 2022.

[29] Giampaolo, et al Bovenzi, "A big data-enabled hierarchical framework for traffic classification," *IEEE Transactions on Network Science and Engineering 7.4 (2020): 2608-2619.*, 2020.

[30] Giuseppe, et al. Aceto, "Mobile encrypted traffic classification using deep learning: Experimental evaluation, lessons learned, and challenges," *IEEE Transactions on Network and Service Management 16.2*, 2019.

[31] Pablo, et al. Belzarena, "SDN-based overlay networks for QoS-aware routing," *Proceedings of the 2016 workshop on Fostering Latin-American Research in Data Communication Networks*, 2016.

[32] Rutvij H., Rui Tan, and Sagar V. Ramani Jhaveri, "Real-time QoS-aware routing scheme in SDN-based robotic cyber-physical systems,"

[32] *2019 IEEE 5th International Conference on Mechatronics System and Robots (ICMSR). IEEE*, 2019.

[33] Noora, and Richard E. Overill Al Khater, "Network traffic classification techniques and challenges," *2015 Tenth international conference on digital information management (ICDIM). IEEE*, 2015.

[34] Omran MA, et al Alssaheli, "Software Defined Network based Load Balancing for Network Performance Evaluation," *International Journal of Advanced Computer Science and Applications*, 2022.

[35] Alfredo, et al. Nascita, "XAI meets mobile traffic classification: Understanding and improving multimodal deep learning architectures," *IEEE Transactions on Network and Service Management 18.4(2021): 4225-4246.*, 2021.

[36] Shi, Yuanjun Xia, and Tao Peng Dong, "Traffic identification model based on generative adversarial deep convolutional network," *Annals of Telecommunication*, 2021.

[37] Ali Malik, ""Intelligent SDN Traffic Classification using Deep Learning"," *Campus Conference Paper*, May 2020.

[38] Yuanjun Xia, and Tao Peng Shi Dong, "Network abnormal traffic detection model based on semi-supervised deep reinforcement learning," *IEEE Transactions on Network and Service Management 18.4*, 2021.

[39] Shi, and Yuanjun Xia Dong, "Network traffic identification in packet sampling environment," *Digital Communications and Networks*, 2022.

[40] Chang, et al. Liu, "Fs-net: A flow sequence network for encrypted traffic classification," *IEEE INFOCOM 2019-IEEE Conference On Computer Communications*, 2019.

[41] Muhammad Basit, et al. Umair, "An Efficient Internet Traffic Classification System Using Deep Learning for IoT," *arXiv preprint arXiv:2107.12193*, 2021.

[42] Shi, and Ruixuan Li Dong, "Traffic identification method based on multiple probabilistic neural network model," *Neural Computing and Applications 31.2*, 2019.

[43] Niloofar, Weston Jackson, and Derrick Liu Bayat, "Deep Learning for Network Traffic Classification," *"arXiv preprint arXiv:2106.12693*, 2021.

[44] R.Thamilselvan K. Tamil Selvi, ""International Journal Of Scientific & Technology Research, Issue 02, February 2020 0 ISSN 2277-8616 2034 IJSTR©2020 www.ijstr.org," *"Deep Learning Based Traffic Classification In Software Defined Networking"*, vol. Volume 9, no. Issue 02, Feburary 2020.

[45] Ren-Hung, et al. Hwang, "An LSTM-based deep learning approach for classifying malicious traffic at the packet level," *Applied Sciences*, 2019.

[46] Manuel, et al. Lopez-Martin, "Conditional variational autoencoder for prediction and feature recovery applied to intrusion detection in iot," *Sensors 17.9 (2017): 1967*, 2017.

[47] Franco Edian F. et al, "Performance comparison of deep learning autoencoders for cancer subtype detection using multi-omics data," *Cancers*, 2021.

[48] Emmanuel, Ioannis E. Livieris, and Panagiotis E. Pintelas Pintelas, "A convolutional autoencoder topology for classification in high-dimensional noisy image datasets," *Sensors 21.22*, 2021.

[49] Qiuyu Zhang, ""A Classification Supervised Auto-Encoder Based on Predefined Evenly-Distributed Class Centroids"," *School of Communication and Information Engineering, Shanghai University, Shanghai 200444, China*, 2019.

[50] R. Alshammari and A.N. Zincir-Heywood, "Can Encrypted Traffic be identified without

Port Numbers, IP Addresses and Payload Inspection?," *Journal of Computer Networks*, 2011.

[51] zgzhengSEU, "nicauca Version2 Dataset," 2021.

[52] Marcus Vinicius Brito da, et al. Silva, "Identifying elephant flows using dynamic thresholds in programmable IXP networks," *Journal of Internet Services and Applications*, 2020.

[53] Tam. G, "Cisco Application Centric Infrastructure," *Americas Headquarters,Cisco Systems, Inc,170 West Tasman Drive,San Jose, CA 95134-1706,USA*, 2018.

[54] Shi Dong and Wei Liu Yongfeng Cui, "Feature Selection Algorithm Based on Correlation between Muti Metric Network Traffic Flow Features," *The International Arab Journal of Information Technology*, vol. Vol. 14, May 2017.

[55] Samira, et al Pouyanfar, "A survey on deep learning: Algorithms, techniques, and applications," *ACM Computing Surveys (CSUR)*, 2018.

[56] P.Amudha and S.Sivakumari Amitha Mathew, "Deep Learning Techniques: An Overview," *International conference on advanced machine learning technologies and applications*, 2020.

[57] Ons, Kandaraj Piamrat, and Benoît Parrein Aouedi, ""Decision tree-based blending method using deep-learning for network management," *NOMS 2022-2022 IEEE/IFIP Network Operations and Management Symposium*, 2022.

[58] Xinyi, Chunxiang Gu, and Fushan Wei. Hu, "CLD-Net: a network combining CNN and LSTM for internet encrypted traffic classification," *Security and Communication Networks* , 2021.

[59] Ons, Kandaraj Piamrat, and Benoît Parrein Aouedi, "Intelligent Traffic Management in Next-Generation Networks," *Future internet*, 2022.

[60] Dario Piga,dynoNet Marco Forgione, "A neural network architecture for learning dynamical systems," *IDSIA Dalle Molle Institute for Artificial Intelligence* , April 2021.

[61] Sachin Malpe, "Automated leaf disease detection and treatment recommendation using Transfer Learning," *National College of Ireland*, 2019. [Online]. http://norma.ncirl.ie/id/eprint/3850

[62] Jeff Bilmes, "Underfitting and Overfitting in Machine Learning," September 26, 2020.

[63] Noam Koenigstein and Raja Giryes Dor Bank, "Autoencoder] 12 Mar ," *arXiv:2003.05991v1 [cs.LG*, 2020.

[64] Hong Xu and Baochun Li, ""TinyFlow: Breaking Elephants Down Into Mice in Data Center Networks"," *Toronto*, TinyFlow2014 IEEE 2014.

[65] E.F. Franco et al., "Performance,Comparison of Deep Learning Autoencoders for Cancer Subtype Detection Using Multi-Omics Data," *Cancers 2021*, 2021.

[66] L. T., et al. Ibrahim, "Online traffic measurement and analysis in big data: Comparative research review ," *Am. J. Appl. Sci 13.4 : 420-431.*, 2016.

[67] Thomas Fischer and Krauss Christopher, "Deep learning with long short-term Memory networks for financial market predictions," *European journal of operational research* , 2018.

[68] Hung Viet, et al. Pham, "Problems and opportunities in training deep learning software systems: An analysis of variance," *Proceedings of the 35th IEEE/ACM international conference on automated software engineering*, 2020.

[69] Kandaraj Piamrat, Salima Hamma Perera Menuka, "Network Traffic Classification using

Machine Learning for Software Defined Networks," *IFIP International Conference on Machine Learning for Networking (MLN'2019)*, 2019.

[70] Dr. Smitha Rao Saahil Afaq, "Significance Of Epochs On Training A Neural Network," *INTERNATIONAL JOURNAL OF SCIENTIFIC & TECHNOLOGY RESEARCH VOLUME 9, ISSUE 06*, 2020.

[71] Khan Muhammad and et al, "Deep Learning for Safe Autonomous Driving: Current Challenges and Future Directions," *IEEE Transactions on Intelligent Transportation Systems PP(99):1-21*, 2020.

[72] Shi. Dong, "Multi class SVM algorithm with active learning for network traffic classification," *Expert Systems with Applications*, 2021.

[73] Binxin Ru and et al., "Speedy Performance Estimation for Neural Architecture Search," *arXiv:2006.04492v2*, 2021.

[74] Marthinus W. Theunissen, Marelie H. Davel Arthur E.W. Venter, "PRE-INTERPOLATION LOSS BEHAVIOUR IN NEURAL NETWORKS," *arXiv:2103.07986v1*, 2021.

[75] YeongHyeon Park, "Concise Logarithmic Loss Function for Robust Training of Anomaly Detection Model," *arXiv preprint arXiv:2201.05748 (2022)*, 2022.

[76] Suhua, et al. Lei, "How training data affect the accuracy and robustness of neural networks for image classification," *How training data affect the accuracy and robustness of neural networks for image classification*, 2018.

[77] Alper, Yeşim Benal Öztekin, and Hüseyin Duran Taner, "Performance analysis of deep learning CNN models for variety classification in hazelnut," *Sustainability 13.12 : 6527.*, 2021.

[78] Quentin, and Daniel Aloise. Fournier, "Empirical comparison between autoencoders and traditional dimensionality reduction methods," *2019 IEEE Second International Conference on Artificial Intelligence and Knowledge Engineering (AIKE). IEE*, 2019.

[79] aptarshi Sengupta and et al, "A Review of Deep Learning with Special Emphasis on Architectures, Applications and Recent Trends," *IEEE Transaction on XXX, VOL. XX, NO. YY.*, 2019.

[80] Pavlo M. Radiuk, "Impact of Training Set Batch Size on the Performance of Convolutional Neural Networks for Diverse Datasets," *Information Technology and Management Scie*, 2017.

[81] A. H., Draper-Gil, G., Mamun, M. S. I., & Ghorbani, A. A. Lashkari, "Characterization of Tor Traffic using Time based Features," *InICISSP*, 2017.

[82] Bushra Mohammed, Usman Humayun, Ahmed Abdelaziz, Suleman Khan, M Akhtar Ali, Muhammad Imraadn and Muhammad Nadzir Marsono Mosab Hamdan, ""Flow-aware elephant flow detection for software-defined networks"," *IEEE Access*, 2020.

[83] Yu-Chiao, et al Chiu, "Predicting drug response of tumors from integrated genomic profiles by deep neural network," *BMC medical genomics 12.1: 143-155*, 2019.

[84] Peter L Dordal, "An Introduction to Computer Networks ," 2020.

[85] Lili Zhu and Petros Spachos, "Towards Image Classification with Machine Learning Methodologies for Smartphones," *University of Guelph, Guelph, ON N1G 2W1*, October 2019.

[86] Fan and R. Liu, "Investigation of machine learning based network traffic classification ," *in 2017 International Symposium on Wireless Communication Systems (ISWCS)*, 2017.